# Stability, electronic structure, and magnetic moment of Vanadium phthalocyanine grafted to the Au(111) surface

Manel Mabrouk [1, 2] and Jacek A. Majewski [1]

[1] University of Warsaw, Faculty of Physics, Pasteura 5, 02-093 Warsaw, Poland

[2] Aix-Marseille Université, CNRS, IM2NP-UMR 7334, 13397 Marseille Cedex 20, France

**Abstract**

The studies of electronic and magnetic properties of V-Pc molecule adsorbed onto Au(111) surface are based on *ab-initio* calculations in the framework of density functional theory. We compute adsorption energies, investigate interaction mechanisms between constituents of the hybrid system consisting of V-Pc molecule and Au surface, and determine geometry changes in the system, particularly in the grafted molecule. We find out that the energetically most stable configuration of the V-Pc/Au(111) occurs when V-Pc is grafted to the Au surface's fcc site, which leads to the reduction of the point group symmetry of the hybrid system in comparison to the free standing V-Pc molecule. Further, our studies reveal that the electronic structure and magnetic properties of the V-Pc change significantly after adsorption to the Au(111). Generally, these studies shed light on physical mechanisms of the V-Pc adsorption to metallic surfaces and open up new prospects for design of novel spintronic devices.

## 1. Introduction

The transition-metal phthalocyanine (TM-Pc) molecules represent a related class of molecular magnets and attracted considerable interest mostly because of the proven possibility to tune their magnetic properties, which opens prospect for achieving of required functionalities and variety of applications [1-4]. Typically, the macrocycles have aromatic character, and therefore, exhibit high thermal and chemical stability. Changing the TM ion in the center, attaching a ligand to the TM ion, and/or attaching different functional groups to the macrocycle can lead to the vast modification of chemical properties of the whole molecule. TM-Pc molecules were mostly employed in organic light-emitting devices [5, 6], sensors [7-10], solar cells [11, 12], catalysts [13-15], and photodynamic therapy [16]. The deep understanding of the interaction between organic molecules and surfaces is not only interesting from the physical point of view but plays very important role for future technological applications. For example, hybrid systems consisting of paramagnetic molecules and metallic substrates might be promising candidates for the miniaturization of magnetic and electric devices, just to avoid size limits of conventional technology.

Previously, TM-phthalocyanines with various transition metals (TM = Cr, Mn, Fe, Co, Ni, Cu, an Zn) on Au(111) surface were studied experimentally as well as theoretically [17-22]. Also there are calculations of tetraphenylporphyrin molecules (belonging to the same family as phthalocyanines) grafted to surfaces of noble metals [20].

Concerning phthalocyanines containing vanadium, one has to point out the theoretical calculations for the three free standing vanadium phthalocyaninato (Pc) based complexes: PcV, PcVO, and PcVI [23], where the XAS spectra were modeled exploring the DFT/ROCIS method. The electronic states and magnetic interactions were investigated experimentally for vanadyl phthalocyanine (VO-Pc) [24] on Si(111)-(7x7) reconstructed surface [25], on Ag(111) surface [25], and also on ferromagnetic Fe, Co, and Ni films [26]. The V-Pc monolayers and multilayers were synthesized on Ag(111) substrates and their electronic and magnetic states were analyzed [27].

Here, we study, not addressed up to now, the hybrid V-Pc/Au(111) system. Our theoretical work should provide physical understanding of the interactions between the two constituents of the hybrid system, and also determine surface induced changes of the magnetic properties of V-Pc.

The paper is organized as follows. In section 2, we describe the employed methodology. In section 3, we present the results of the studies and their discussion. Finally, in section 4, the paper is concluded.

## 2. Computational methods

In our studies, we employ spin polarized Kohn-Sham realization of the density functional theory as implemented in the Vienna *ab-initio* Simulation Package (VASP) [28]. The projector augmented wave (PAW) [29, 30] pseudo-potentials are employed to account for the electron-ion interactions. The electron-electron interactions are accounted for within the spin-polarized generalized gradient (GGA) exchange-correlation functional proposed by Perdew and Wang (PW91) [31, 32], therefore, hereafter the applied computational scheme is dubbed as SGGA. The Au(111) substrate is modeled using periodic supercells of four layers with (7× 8) lateral unit cell. The calculated lattice constant of the slab is equal to 4.17 Å. To avoid spurious interaction between the surfaces along the z-axis, a vacuum layer of 16.85 Å is placed into the supercell between the slabs. The V-Pc molecule is placed in the supercell in such a way that its plane is parallel to the Au surface. In this case, the supercell contains altogether 281 atoms. In the calculations of the hybrid system, the molecule's geometry is fully relaxed and the Au(111) layers are kept rigid. The superlattice geometry optimization is performed until the residual forces are smaller than 0.02 eV/Å. A plane-wave cutoff energy of 400 eV is employed, and the Brillouin zone is sampled only in the Γ-point, which is justified by the rather large dimensions of the employed supercell. Altogether, the performed convergence tests indicate that these settings guarantee the precision of the total energy calculations up to 0.01eV (within the chosen model). For test calculations (performed only for free V-Pc molecule), we have used also SGGA+$U$ approach with the Hubbard-like Coulomb correction terms in the d shell of the transition metal, as introduced some time ago by Liechtenstein *et al.* [33], using the value of $U$ equal to 5 eV, as proposed earlier in the literature [34, 35]. Although typical approaches to the calculations of adsorption of organic molecules to metallic surfaces do not include dispersive forces, we have performed some test calculations employing DFT+D3 model of van der Waals (vdW) interactions proposed by Grimme [36]. This allows us to assess the role of dispersive forces for the adhesion of V-Pc molecule to the Au(111) metallic surface.

## 3. Results and discussion

Here, we would like to present the bonding mechanism of the V-Pc molecule to the gold surface emerging from the performed calculations. We describe details of the geometry, magnetic properties, and energetics of the hybrid system.

The adsorption energy $E_a$ of the V-Pc molecule to the Au surface is calculated via the following formula:

$$E_a = E_{system} - E_{V-Pc} - E_{surface}; \quad (Eq. 1)$$

where $E_{system}, E_{V-Pc}$, and $E_{surface}$ are the total energies of the whole hybrid V-Pc/Au(111) system, the isolated V-Pc molecule, and the Au(111) surface modeled by a four layer slab, respectively. All three energies in Eq. 1 are computed employing identical supercells with the same kinetic energy cutoffs. We have calculated adsorption energy for V-Pc molecule placed in a series of possible positions on the Au surface. This allows us to determine the energetically most favorable adsorption site, and further the equilibrium geometry of the hybrid system. Before we turn to the discussion of the adhesion of the V-Pc molecule to the Au surface, let us summarize shortly the results for free standing V-Pc.

**3.1 Free standing V-Pc molecule**

The planar V-Pc molecule is shown schematically in Figure 1. The central V atom is surrounded by four pyrrole rings (which are bridged by four additional N atoms) with benzene rings attached to them. The V-Pc molecule exhibits $D_{4h}$ symmetry.

We have performed SGGA calculations for V-Pc molecule and determined it's geometry, magnetic moments, total energy, and electronic structure. All these quantities are shown in Table 1. The calculated distance between V and N atoms d(V-N) equals to 1.99 Å and compares very well with previous calculations predicting the distance d(V-N) to be equal to 1.99 Å [34, 23]. The total magnetic moment of the whole molecule M is equal to 3.00 $\mu_B$ and compares fairly well with magnetic moment computed for V-Pc polymer [34]. The magnetic moment contributed by the 3d states of the central transition metal $M_d$ is equal to 2.14 $\mu_B$ and is minimally smaller than the magnetic moment of the whole V atom $M_m$. In the $D_{4h}$ symmetry, the five-fold degeneracy of the 3d atomic levels is removed, and d-states manifold is splitting into four levels, not degenerated $d_{x^2-y^2}$, $d_{z^2}$, $d_{xy}$ and doubly degenerated $d_{yz}$ and $d_{xz}$ states. This is illustrated in Figure 2, where also the splitting of the d-states in $O_h$ symmetry is given for comparison. The 3d levels of Vanadium atom in V-Pc follow this splitting pattern as can be clearly seen in Figure 3, where the projected on d-states density of states (PDOS) for the free standing molecule is depicted. The total density of states (DOS) of the V-Pc molecule is shown in Figure 4. It is clearly seen that the spin polarization of the molecule originates mostly from the d-states with energies around Fermi energy, and also that the HOMO-LUMO gap is present in V-Pc.

It is well known that approximate exchange-correlation functionals (such as used here SGGA) lead to spurious self-interaction and, in results, following artificial delocalization of the d-states. As remedy to this problem, the so-called DFT+*U* scheme is sometimes employed for systems, where d-states play an important role [34-35, 37-39]. Therefore, we have also performed calculations within *U* corrected DFT scheme, which we name SGGA+*U*. The results of these calculations are given in the lowest row of Table 1. As expected the local magnetic moment originating from the d-states in now slightly higher, 2.43 $\mu_B$ in comparison to 2.14 $\mu_B$ obtained without *U* correction. The other quantities do not change significantly, which allows us to believe that the representative results of V-Pc adsorption to the Au surface can be achieved with less computationally demanding SGGA scheme.

**3.2 V-Pc molecule grafted to the Au(111) surface**

The essential part of the calculations for the hybrid system of V-Pc molecule grafted to the Au (111) surface is determination of the energetically most stable geometry. We consider seven possible placements of the V-Pc molecule on the Au(111) surface, which are depicted in Figure 5 and indicated as: top-I (T-I), top-II (T-II),top-III (T-III), bridge-I (B-I), bridge-II (B-II), hexagonal closed packed (hcp), and cubic closed packed (fcc). For each of these starting positions, we perform optimization of the geometry and find the energy of the hybrid systems employing the SGGA computational scheme. The calculated total energies of the free-standing V-Pc molecule and the Au slab, with the same supercell employed as for calculations of the hybrid system, are equal to -424.150 eV and -679.686 eV, respectively. Then subtracting these energies from the total energies of the hybrid system (Equation 1), we obtain adsorption energies $E_a$ of the V-Pc molecule grafted to the Au(111) surface. For the considered adsorption sites T-I, T-II, T-III, B-I, B-II, bcc, and fcc the adsorption energies are equal to -1.931, -1.988, -1.988, -1.993, -1.96,-2.022, and -2.05 eV, respectively. As one can see, the hybrid system exhibits rather strong adhesion forces between the V-Pc molecule and the Au surface, which are only slightly dependent on the V-Pc grafting position. Nevertheless, the energetically most favorable adsorption position turns out to be the fcc one. The results of calculations are summarized in Table 2. The optimized distances between central phthalocyanine atom V and Au(111) surface d(V-Au) for all 7 grafting positions of the V-Pc molecule are depicted in the third column of the Table 2. As intuitively expected, the shortest distance, d(V-Au) = 2.51 Å, is obtained for molecule grafting position exhibiting the strongest adsorption (fcc). Generally, the distances d(V-Au) for other grafting position follow the trend for the adsorption energy, the stronger adsorption the smaller distance to the surface. The adsorption of the V-Pc molecule to the Au(111) surface causes considerable changes of the geometry of the V-Pc adsorbate, as it is illustrated in panel (h) of the Figure 5, for the energetically most stable configuration of the V-Pc/Au(111) system. First, for majority of the considered grafting positions, the symmetry of the hybrid system is lower than $\mathbf{D_{4h}}$ characterizing free standing V-Pc molecule. In result of this tetragonal symmetry breaking, the distances between central V atom and its nearest N neighbors d(V-N) slightly differ. They vary in the range of 2.02−2.05 Å, and they are also considerably larger than the V-N distances in the free standing V-Pc molecule. Second, the V-Pc molecule gets slightly bent, with V atom moving towards the Au atoms and the surrounding skeleton moving apart from the Au surface. We ascribe this effect of geometry distortion of the V-Pc adsorbed molecule to the electronic charge transfer from the V-Pc to the Au(111) metallic surface. These charge transfer effect is clearly seen when one analyzes the local magnetic moments in the V-Pc/Au(111) hybrid system. Whereas the total magnetic moment of the hybrid system M (see column 4 in Figure 5) diminishes by 0.3−0.4 $\mu_B$ in comparison to the case of free standing V-Pc molecule (where M = 3.0 $\mu_B$), the magnetic moment on the V atom $M_m$ (see column 5 in Figure 5) decreases by 0.55−0.69 $\mu_B$, and the local moment of the 3d vanadium electrons $M_d$ (see column 6 in Figure 5) gets smaller by 0.39−0.54 $\mu_B$.

Finally, we present electronic structure of the hybrid V-Pc/Au(111) system with V-Pc molecule adsorbed at equilibrium fcc surface site. The partial and total densities of states (PDOS& DOS) for the most energetically favorable configuration of hybrid system are shown in Figures 6 and 7 and can be compared with results for the free V-Pc molecule presented in Figures 3 & 4. The Fermi energy $E_F$ is set to zero in all the figures. The interaction between the V-Pc molecule and the Au(111) substrate can evidently affect the charge distribution in the constituents of the hybrid system, and further influence the geometry of the V-Pc molecule at the interface and its binding strength to the Au surface. As can be seen at the figures, the adhesion of V-Pc to the Au surface mostly indices strong changes in the d-states originating from the central V atom. Redistribution of electrons over the d states causes that the band gap disappears (see Figures 4 and 5) and the hybrid system exhibits a metallic character for both spin-up and spin-down bands. One can also observe shits of peaks and change of their intensities. However, the $d_{x^2-y^2}$ orbital remains empty inthe hybrid system, and the Fermi energy cuts the peaks associated with the $d_{xz}$, $d_{yz}$ and $d_{z^2}$. All these findings presented above strongly suggest that the main mechanism of bonding the V-Pc to the Au(111) surface is achieved through the banding of central vanadium atom to the Au surface.

However, in the hybrid system, one observes also the changes of the PDOS for 2p states of nitrogen (N1) and carbon (C1) atoms of the V-Pc molecule depicted in Figure 8, where N1 and C1 atoms are related to the pyrrole rings and are the nearest neighbors and the second nearest neighbors of V atom, respectively (see Figure 1). Therefore, the local changes of the electronic charge on these C and N atoms, also gives contribution to the binding of the V-Pc molecule to the Au surface.

## 3.3 Chemisorption vs. physisorption character of the V-Pc adhesion to the Au (111) surface

The long standing problem of the physico-chemistry of the transition metal phthalocyanines on metallic surfaces is the issue of the character of the adsorption, i.e., chemisorption vs. physisorption. The results described in the section 3 have been obtained employing PW exchange-correlation density functional. As it is well known [40], the PW functional is not able to account for dispersive forces. However, without taking van der Waals interactions into account, our studies demonstrate that the V-Pc molecule adsorbs fairly strongly to the Au (111) surface, practically in all considered adsorption sites, just indicating the existence of the covalent binding forces. The adhesion happens mostly through the binding of the transition metal atom (V in our case) to the Au surface.

To assess the role of the dispersive forces in the adsorption mechanisms of the V-Pc to the Au surface, we have performed the calculations employing the simple DFT-D3 scheme of Grimme [36], where the pairwise corrections proportional to $R_{ij}^{-6}$ are added to $R_{ij}^{-1}$ ion-ion repulsive term. As it was found in Ref. [40],such simple interatomic corrections when added to the large class of functionals provide for satisfactory level of accuracy irrespectively of the underlying functionals. This has been also confirmed in the very recent extensive studies of the various van der Waals forces schemes within the framework of the DFT [41]. We have performed DFT+D3 the calculations only for the V-Pc molecule adsorbed to the Au(111) surface at the fcc site (the most stable configuration according to the calculation with PW functional alone). The most pronounced effect of the employment of the DFT+D3 scheme is that the adsorption energy of the V-Pc to the Au surface changes from -2.05 eV (for PBE) to -5.42 eV. The nearly entire effect of the van der Waals forces on the adsorption energy comes from the total energy of the hybrid V-Pc/Au(111) system. The van der Waals corrections lower the total energies of all three components of the Eqn. (1) for the adsorption energy, however, the energy of the free standing V-Pc molecule is lowered by -1.06 eV, and the energy of the Au slab is lowered only by -0.15 eV. The vdW correction has practically no effect on the distance between central vanadium atom and the nearest N neighbors (being now equal to 2.00 Å) and on the magnetic moments of the V-Pc molecule (for example, the total magnetic moment of the V-Pc decreases by 0.02 $\mu_B$). In spite of the fact that the vdW interactions lower the adsorption energy of the V-Pc on Au surface by -3.37 eV, the distance between the Au(111) surface and the V atom is larger by 0.11 Å, when the DFT+D3 computational scheme is employed. The DFT+D3 calculations lead also to the total magnetic moment of the hybrid system equal to 1.88 $\mu_B$ (lower by 0.71 $\mu_B$ in comparison to the PW result).

The similar effects of the vdW interaction on adsorption mechanism have been previously observed for Co-Pc/Ag(111) hybrid system [42], studied within DFT-D2 scheme [43] that minimally differs from the DFT-D3 one, and also in the paper dealing with the vdW forces in bilayer systems consisting of graphene coupled to SiC or h-BN [44]. Definitely, in the cases of V-Pc/Au(111) and Co-Pc/Ag(111), the vdW interaction between transition metal phthalocyanines and metallic surfaces makes the adsorption stronger, however, we agree with the suggestion of the authors of reference [42] that the role of vdW forces may be somewhat overestimated in the simple DFT+D3 computational scheme.

## 4. Conclusions

In this paper, we have presented the results of theoretical investigation of V-Pc adsorbed on Au(111) substrate performed at the DFT level of theory. This study sheds light on the physico-chemical mechanisms of the adhesion of V-Pc molecule to the metallic surface. Even when the dispersive forces are not taken into account, the studied systems exhibits fairly strong tendency to the adsorption of the V-Pc molecule to the Au surface, which is only weakly dependent on the adsorption site. This chemisorption is mostly driven by the bonding of the 3d-shell V atom to the Au surface. The free standing V-Pc molecule exhibits HOMO-LUMO gap, however the hybrid systems gets metallic after the adsorption of V-Pc to the Au(111) surface. The redistribution of the electronic charge caused by the V-Pc adsorption induces small changes of the V-Pc geometry and diminishes its magnetic moment. The dispersive van der Waals interactions (treated in the paper within the DFT-D3 computational scheme) strongly increase the adsorption of the V-Pc molecule to the Au surface, however, this effect might be exaggerated. The vdW interaction between V-Pc and Au(111) changes also slightly the geometry of the adsorbed V-Pc and significantly reduces the total magnetic moment of the system. We believe that the present theoretical results provide an important insight into the physico-chemistry of the transition metal phthalocyanines on metallic substrates and could facilitate design of future spintronics devices.

**Acknowledgments**

This work has been supported by National Science Centre Poland (UMO 2016/23/B/ST3/03567) and by the Centre Informatique National de l'Enseignement Supérieur (CINES), Project No. c2016096873.

**Competing financial interests**

The authors declare no competing financial interests.

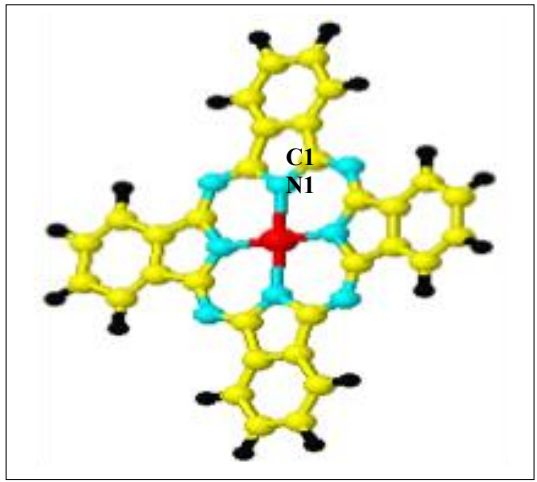

**Figure 1**. The planar V-Pc molecule (Red ball indicates the central vanadium atom, cyan balls-N atoms, yellow balls-C atoms, and black balls-H atoms) exhibiting the $D_{4h}$ symmetry with symmetry axis going through the central atom. The nearest neighbor shell to the V atom consists of 4 nitrogen atoms (exemplary one indicated by N1), whereas the second neighbor shell is built of C atoms (a representant indicated by C1).

**Table 1**. Theoretical results for free standing V-Pc molecule obtained within SGGA and SGGA+$U$ computational schemes. The d(V-N) indicates the distance between central V atom and the nearest neighbor N atoms. The total magnetic moment of the molecule M, the local magnetic moment on the central V atom ($M_m$), and the local magnetic moment on the central atom originating from the 3d states ($M_d$) are also depicted.

| Method | d(V-N) (Å) | M ($\mu_B$) | $M_m$ ($\mu_B$) | $M_d$ ($\mu_B$) |
|---|---|---|---|---|
| SGGA | 1.99 | 3.00 | 2.24 | 2.14 |
| SGGA+$U$ | 2.01 | 3.00 | 2.55 | 2.43 |

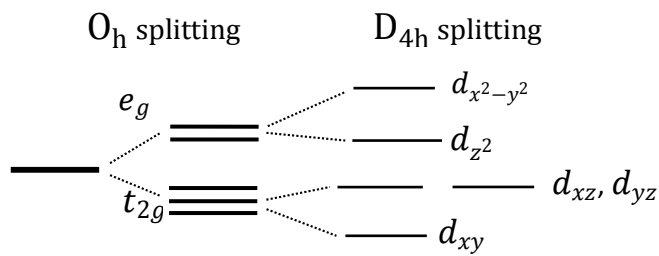

**Figure 2**. Schematic diagram of 3d level splitting in the square planar symmetry ($D_{4h}$).

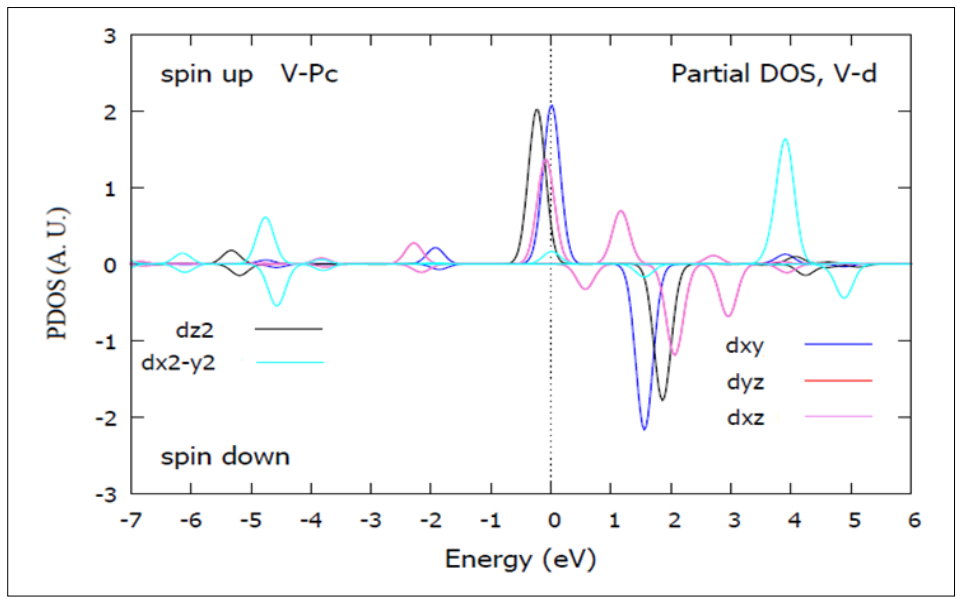

**Figure 3**. Spin resolved 3d states projected density of states of V-Pc molecule. The $d_{yz}$ and $d_{xz}$ orbitals are degenerated, so we cannot distinguish between them. The Fermi energy $E_F$ has been set to zero and indicated by the dotted line.

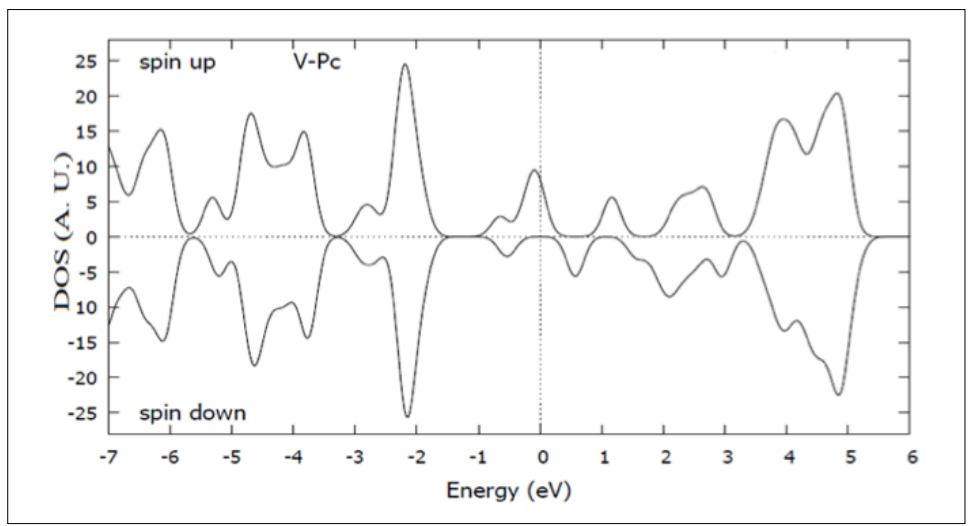

**Figure 4**. Spin resolved total DOS of the V-Pc molecule. The Fermi energy $E_F$ is set to zero. The spin polarization originating from states with energies around Fermi energy is clearly seen. The HOMO-LUMO gap can be also noticed.

**Table 2.** The calculated quantities for the hybrid system constituted of the V-Pc molecule adsorbed on various positions at the Au (111) surface. The positions listed in the first column of the table have been described in the text and in details presented in Figure 5. The listed quantities are indicated as follows: $E_a$ (eV) the adsorption energy; d(V-Au) (Å) the distance between Au surface and V atom; total magnetic moment of the hybrid system M ($\mu_B$); local magnetic moment at V atom $M_m$ ($\mu_B$); local magnetic moment of the d orbitals at the V atom $M_d$ ($\mu_B$); ΔE (meV) adsorption energy of the hybrid systems with the V-Pc molecule attached to a certain place on the Au (111) surface relatively to the adsorption energy of the energetically most stable adsorption position(fcc site), and d(V-N) (Å) the distances between central V atom and surrounding N atoms. Note, that at certain adsorption positions of the V-Pc molecule, the symmetry of the whole hybrid structure is lower than $D_{4h}$, and then all different distances between V and the neighboring N atoms are shown.

| Position | $E_a$ (eV) | d(V-Au) (Å) | M ($\mu_B$) | $M_m$ ($\mu_B$) | $M_d$ ($\mu_B$) | ΔE (meV) | d(V-N) (Å) |
|---|---|---|---|---|---|---|---|
| Top-I | -1.931 | 2.69 | 2.69 | 1.62 | 1.57 | 120 | 2.02/2.03 |
| Top-II | -1.988 | 2.69 | 2.69 | 1.64 | 1.58 | 63 | 2.02/2.03 |
| Top-III | -1.988 | 2.67 | 2.69 | 1.60 | 1.55 | 63 | 2.02 |
| Bridge-I | -1.993 | 2.52 | 2.66 | 1.65 | 1.60 | 58 | 2.03/2.04/2.05 |
| Bridge-II | -1.960 | 2.52 | 2.70 | 1.73 | 1.67 | 91 | 2.03/2.04 |
| Hcp | -2.022 | 2.65 | 2.66 | 1.72 | 1.67 | 29 | 2.03/2.04/2.05 |
| Fcc | -2.051 | 2.51 | 2.59 | 1.75 | 1.69 | 0 | 2.03/2.04/2.05 |

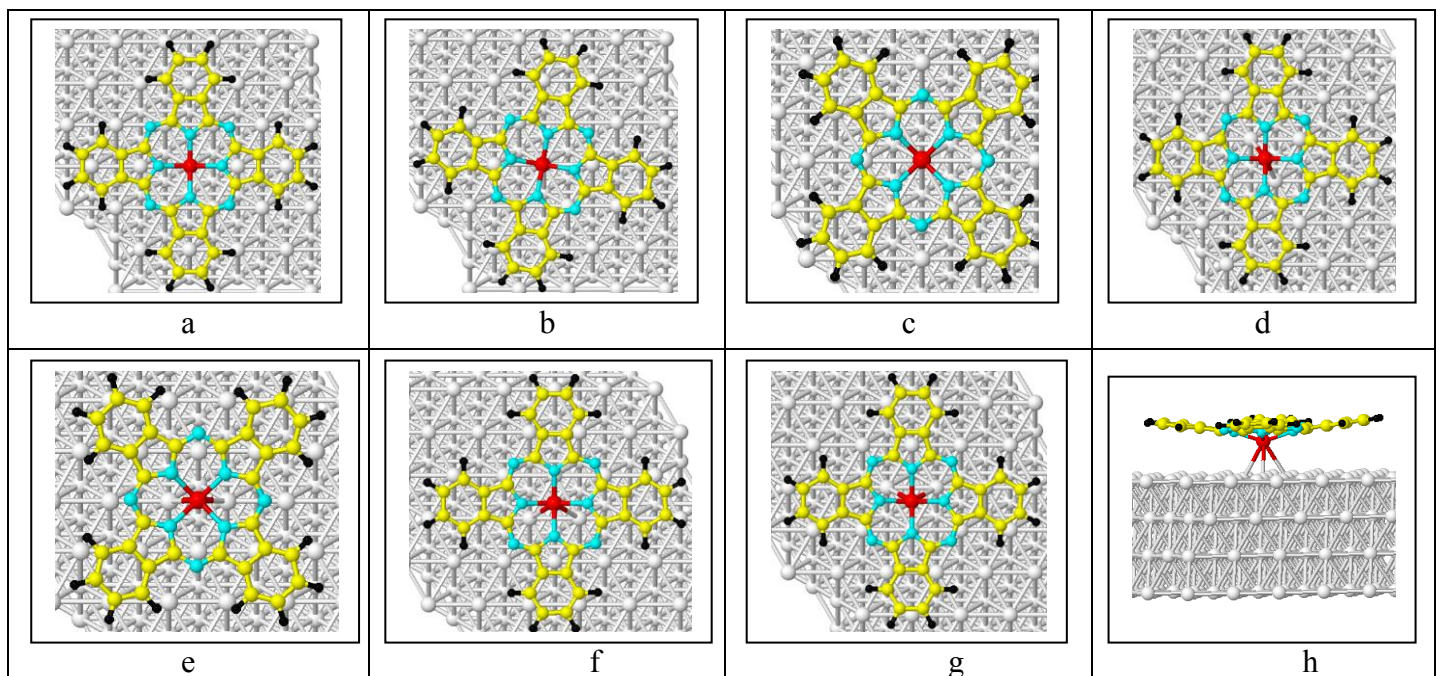

**Figure 5**. Adsorption geometries of the V-Pc molecule on Au(111) surface considered in the paper. (a) Top- I (T-I), (b) T-II, (c) T-III, (d) Bridge- I (B-I), (e) B-II, (f) Hcp, (g) Fcc, and (h) side view of the Fcc site (the energetically most stable position). The V, N, C, H, and Au atoms are indicated by red, cyan, yellow, black, and white balls, respectively.

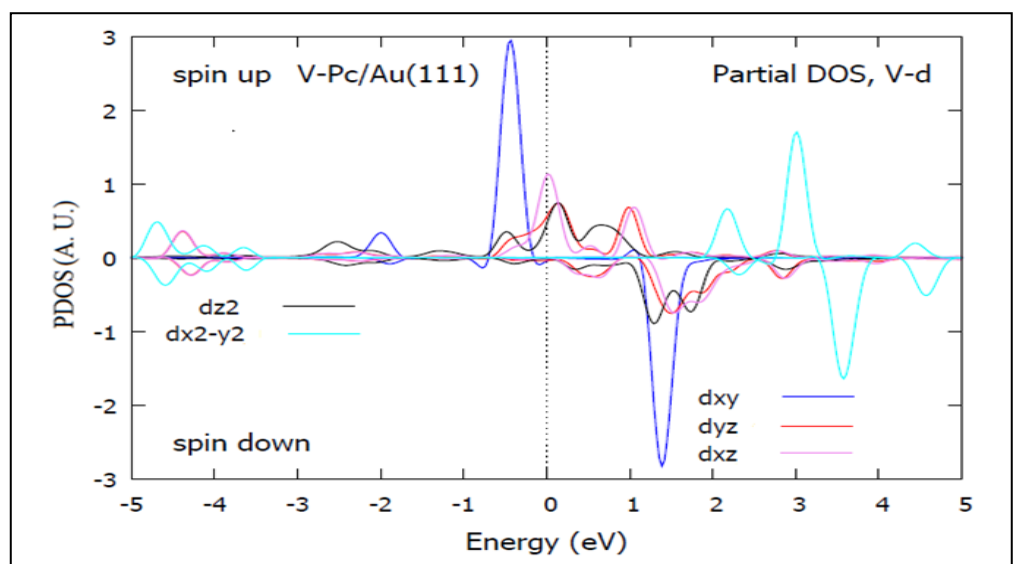

**Figure 6**. PDOS of the V-3d orbitals in the V-Pc/Au(111) hybrid system. The Fermi energy is set to zero.

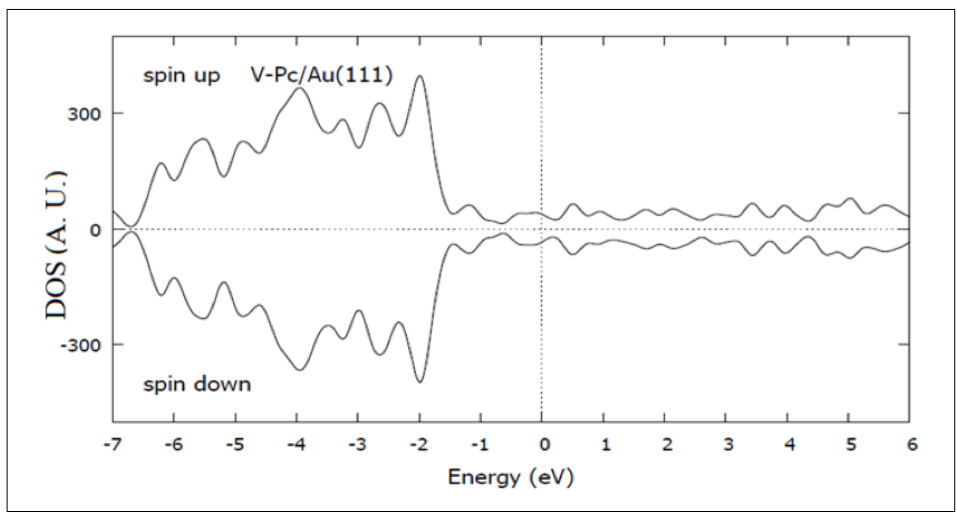

**Figure 7**. DOS for the V-Pc/Au(111) hybrid system. The Fermi energy is set to zero. The metallic character of spin-down and spin-up bands of this system is clearly visible.

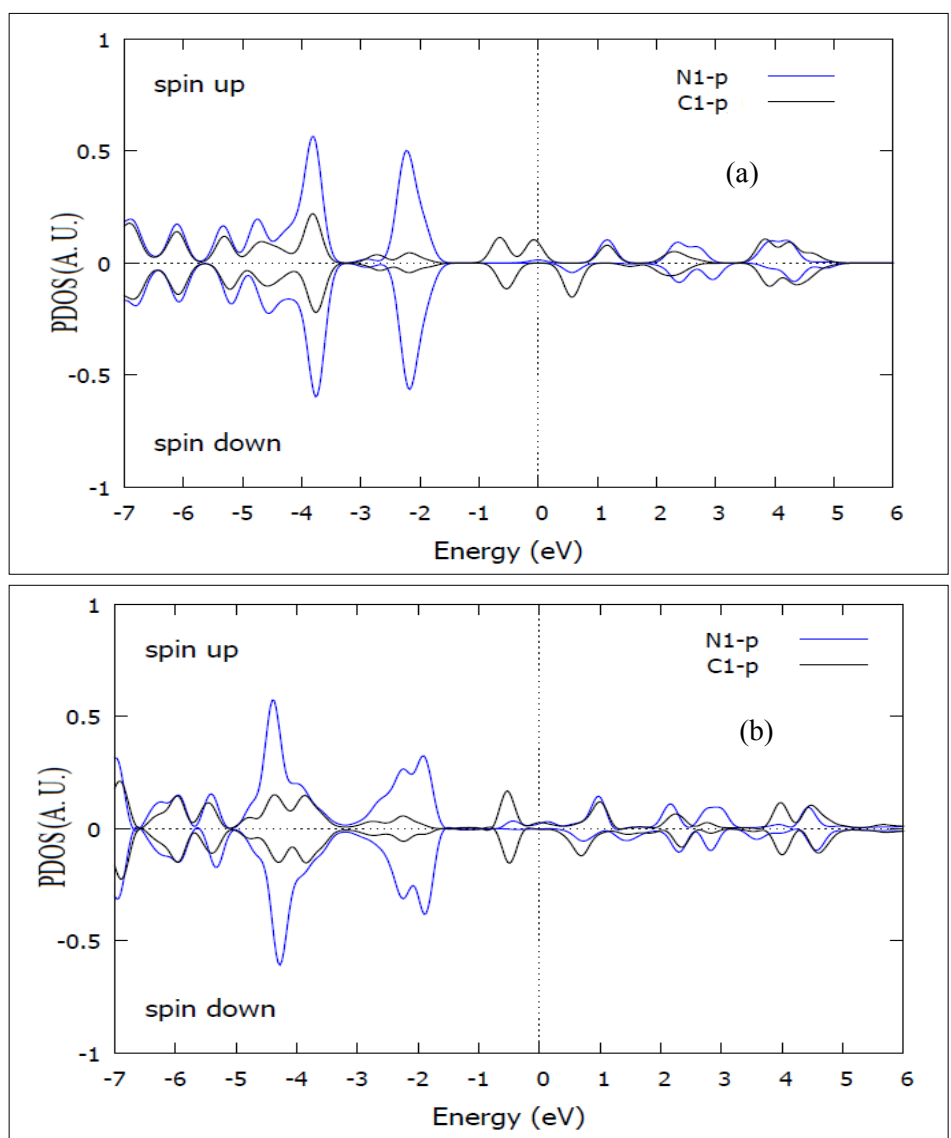

**Figure 8**. The PDOS for 2p states of nitrogen (N1) and carbon (C1) atoms: (a) in the free standing V-Pc molecule, and (b) in the V-Pc/Au(111) hybrid system with V-Pc adsorbed at fcc site of Au surface. The N1 atoms are the nearest neighbors of the central V atom, while the C1 atoms are from the pyrrole rings (see Fig. 1).

17